\documentstyle[12pt]{article}
\renewcommand
\baselinestretch{1.3}

\def\beq{\begin{equation}}
\def\brr{\begin{array}}
\def\err{\end{array}}
\def\eeq{\end{equation}}
\def\bea{\begin{eqnarray}}
\def\eea{\end{eqnarray}}
\def\bs{\bigskip}
\def\tr{\mbox{Tr}\, }
\def\ni{\noindent}
\def\wt{\widetilde}

\def\nn{\nonumber}
\def\ms{\medskip}

\def\dsp{\displaystyle}

\begin{document}

\hfill HUPD-9304

\hfill UB-ECM-PF 93/2

\hfill February 1993

\vspace*{3mm}

\begin{center}

{\LARGE \bf
Renormalization-group improved effective potential for gauge
theories
in curved spacetime}

\vspace{4mm}

\renewcommand
\baselinestretch{0.8}
{\sc E. Elizalde}\footnote{E-mail address: eli @ ebubecm1.bitnet}
\\
{\it Department E.C.M., Faculty of Physics, University of
Barcelona, \\
Diagonal 647, 08028 Barcelona, Spain} \\  and \\
{\sc S.D. Odintsov}\footnote{On  leave from
Tomsk Pedagogical Institute, 634041 Tomsk, Russia. E-mail address:
odintsov @ theo.phys.sci.hiroshima-u.ac.jp} \\ {\it Department of
Physics, Faculty of Science, Hiroshima University, \\
Higashi-Hiroshima 724, Japan}
\ms

\renewcommand
\baselinestretch{1.4}

\vspace{5mm}

{\bf Abstract}

\end{center}

The renormalization-group improved effective potential for an
arbitrary
renormalizable massless gauge theory in curved spacetime is found,
thus
generalizing Coleman-Weinberg's approach corresponding to flat
space.
Some explicit examples are considered, among them: $\lambda
\varphi^4$
theory, scalar electrodynamics, the asymptotically-free SU(2) gauge
model, and the SU(5) GUT theory. The possibility of
curvature-induced
phase transitions is analyzed. It is shown that such a phase
transition
may take place in  a SU(5) inflationary universe. The inclussion of
quantum gravity effects is briefly discussed.

\vspace{5mm}

%\noindent PACS: 04.50, 03.70, 11.17.

\newpage

\ni {\bf 1. Introduction.}
It is well known in modern gauge theories that symmetry breaking
and/or restoration can be caused by some external conditions, like
temperature, external electric and magnetic fields, finite density,
etc. Moreover, these changes in the vacuum structure (i.e.,
symmetry breaking and restoration) may be described as some phase
transition (using the effective potential formalism). Such phase
transitions are extremely important in early universe cosmology.
Specifically, some models of inflationary universe (see [1,2] for
a review and list of references) are based on the first-order phase
transitions which take place during the reheating of the universe
in the grand unification epoch.

However, it is clear that curved spacetime effects in the early
universe (the GUT epoch) cannot be considered at all to be negligibly
small. Therefore, GUTs corresponding to the very early universe
ought to be treated as quantum field theories in {\it curved}
spacetime (for a general review see [3]).

Unfortunately, at present we do not have a clear prescription how
to combine quantum field theory at non-zero temperature and quantum
field theory in curved spacetime (external temperature {\it and}
external gravitational field). In such a situation, it seems
reasonable to investigate in depth the two topics involved: field
theory in curved space and field theory at non-zero temperature,
always with the aim of combining them, later on.
We will concentrate on the first topic here.

In the present paper, we shall study the effective potential of
massless gauge theories in curved spacetime. We will work in the
linear curvature approximation because, as it has been argued, at
least linear curvature terms should be taken into account in the
discussion of the effective potential corresponding to GUTs in the
early universe. Also, quantum corrections with account to gravity
effects should be even more important in a chaotic inflationary
model[1].

By generalizing the Coleman-Weinberg approach [4] (for a review,
see [5]) corresponding to the case of the effective potential in
flat spacetime, at a first instance, we find the renormalization
group (RG) improved effective potential in curved space. Hence, we
extend the one-loop effective potential in curved spacetime, taking
into account all logarithmic corrections. (The one-loop effective
potential in curved spacetime has been calculated previously for
the cases of scalar electrodynamics [6] and arbitrary massless
gauge theory [7]. In these papers the corresponding
curvature-induced phase transitions have been also discussed.)

We shall here present the explicit form of the RG improved
effective potential for $\lambda \varphi^4$, scalar
electrodynamics, the SU(2) gauge model, and SU(5) GUT. The
possibility of corresponding curvature-induced phase transitions will be
briefly discussed. \ms

\ni {\bf 2. RG improved effective potential.}
Let us consider an arbitrary, renormalizable, massless gauge theory
including scalars $\varphi$, spinors $\psi$, and vectors $A_\mu$.
Denote by $\wt{g} \equiv (g,\lambda,h)$ the set of all coupling
constants of the theory ($g$ is the Yang-Mills, $\lambda$ the
scalar, and $h$ the Yukawa coupling).
The tree level potential reads
\beq
V^{(0)}= a\lambda \varphi^4-b\xi R \varphi^2,
\eeq
where $a$ and $b$ are some positive constants, $\xi$ the conformal
coupling [3], and $R$ the scalar curvature.

The RG equation for the effective potential in curved spacetime has
the following form [3,7]:
\beq
\left( \mu \frac{\partial}{\partial \mu} +\beta_{\wt{g}}
\frac{\partial}{\partial \wt{g}} +\delta  \frac{\partial}{\partial
\alpha} + \beta_\xi \frac{\partial}{\partial \xi}
-\gamma \varphi \frac{\partial}{\partial \varphi}  \right)
V=0,
\eeq
where $\alpha$ is the gauge parameter. This RG equation is standard
[4,5] but not for the term connected with $\xi$.

We now split $V$ into $V\equiv V_1 +V_2 \equiv af_1(p,\varphi,\mu)
\varphi^4 -b f_2(p,\varphi,\mu)R \varphi^2$, where  $f_1$ and $f_2$
are some unknown functions and $p=\{ \wt{g}, \alpha, \xi \}$. We
also assume that both $V_1$ and $V_2$ satisfy the RG equation (2).
(Notice that this is in fact a restriction, because in general only
$V$ should satisfy (2), and not $V_1$ and $V_2$ separately.) \, We
choose the Landau gauge for the gauge fields, so that in the
one-loop approximation $\delta =0$ in (2). The other point is that
$\varphi^2 >> |R|$, otherwise we cannot meaningfully expand the
effective potential with an accuracy up to linear curvature terms.

With all these considerations in mind, we can now solve the RG
equation (2) in the way
\beq
V= a\lambda (t) f^4(t) \varphi^4-b\xi (t) f^2(t) R \varphi^2,
\eeq
where $f(t)= \exp \left[ -\int_0^t dt' \, \bar{\gamma} \left(\wt{g}
(t'), \alpha (t'), \xi (t') \right) \right]$, $t=\frac{1}{2} \ln
(\varphi^2/\mu^2)$, $\dot{\wt{g}} (t) =\bar{\beta}_{\wt{g}} (t)$,
$\dot{\alpha} (t) =\bar{\delta} (t)$, $ \dot{\xi} (t)
=\bar{\beta}_\xi (t)$, $\wt{g} (0) =\wt{g}$, $\alpha (0)=\alpha$,
$\xi (0) =\xi$, and $(\bar{\beta}_{\wt{g}}, \bar{\delta},
\bar{\beta}_\xi, \bar{\gamma})$ $=\frac{1}{1+\gamma} (\beta_g,
\delta, \beta_\xi, \gamma)$. Notice that in the solution of Eq. (2)
we have used the following initial conditions:
\beq
V_1(t=0) = a\lambda \varphi^4, \ \ \ \ V_2 (t=0)=-b\xi R \varphi^2.
\eeq
These initial conditions are only slightly different from the ones
used in Refs. [4,7], while that for $V_1$ is the same as in Ref.
[5]. The differences here will lead to some differences in the
non-logarithmic $\varphi^4$ and $R\varphi^2$ terms.

In the one-loop approximation (in which we actually work in this
paper), Eq. (3) remains formally the same, but now with
\beq
f(t)=\exp \left[ -\int_0^t dt' \, \gamma \left(\wt{g} (t'),  \xi
(t')\right)\right], \ \ \dot{\wt{g}} (t) =\beta_{\wt{g}} (t), \ \
\dot{\xi} (t) =\beta_\xi (t), \ \ \wt{g} (0) =\wt{g}, \ \ \xi
(0) =\xi.
\eeq
Expression (5) constitutes our main result for the one-loop RG
improved potential, and can be applied to a variety of gauge
theories.

Let us now consider some examples.
\ms

\ni {\bf (a) $\lambda \varphi^4$ theory.} In this case $\gamma=0$,
and $\beta_\lambda$ and $\beta_\xi$ are well known (see for example
[3,4]); we get from (4)\footnote{Note that the one-loop effective
potential for the $\lambda \Phi^4$ theory has been calculated in Refs.
[16] for different specific background spaces.}
\beq
 V= \frac{\lambda \varphi^4}{4! \left(1-\dsp \frac{3\lambda t}{(4
\pi)^2}\right)} - \frac{1}{2} R\varphi^2 \left[ \frac{1}{6} +
\left( \xi - \frac{1}{6} \right) \left( 1- \frac{3\lambda t}{(4
\pi)^2}\right)^{-1/3} \right],
\eeq
where $t=\frac{1}{2} \ln (\varphi^2/\mu^2)$. In the limit in which
both $\lambda$ and $\lambda t$ are small, our result agrees with
the previously obtained one-loop potential [6,7]. However, as
has been discussed in [4,5] for the case of flat space, Eq. (5) is
actually valid for all $t$ for which the potential does not diverge
(in particular, for any negative value of $t$).
\ms

\ni {\bf (b)  Scalar electrodynamics.}  Using the well-known
results [4,6,7]
\[
\beta_\lambda = \frac{1}{4\pi^2} \left( \frac{5}{6} \lambda^2 -
3e^2\lambda +9e^4 \right), \ \
\beta_\xi = \frac{\xi- \frac{1}{6}}{4\pi^2} \left( \frac{1}{3}
\lambda   -\frac{3}{2}e^2 \right), \ \  \gamma =-
\frac{3e^2}{16\pi^2},
\]
where $\gamma$ is given in the Landau gauge, we obtain
\bea
V&=& \frac{1}{4!} f^4(t) \varphi^4 \left\{ \frac{1}{10} e^2(t)
\left[ \sqrt{719}\, \tan \left( \frac{1}{2}  \sqrt{719}\, \ln e^2(t) +
\theta \right) +19 \right] \right\} \nn \\
&-& \frac{1}{2} R \varphi^2 f^2(t) \left[ \frac{1}{6} + \left( \xi
- \frac{1}{6} \right) \left( 1- \frac{e^2t}{24 \pi^2}\right)^{-
26/5} \cos^{2/5} \left(  \frac{1}{2}  \sqrt{719}\, \ln e^2(t) +
\theta \right) \right. \nn \\
&\cdot& \left. \cos^{-2/5} \left(  \frac{1}{2}  \sqrt{719}\, \ln e^2
 + \theta \right) \right].
\eea
Here $e^2(t)=e^2 \left( 1- \frac{e^2t}{24 \pi^2}\right)^{-1}$,
$\theta$ is an integration constant, which should be chosen such
that $\lambda (t)=\lambda$ when $e^2(t)=e^2$, and $f(t)=
\left( 1- \frac{e^2t}{24 \pi^2}\right)^{-9/2}$. It is interesting
to notice that for very small variations in $e^2(t)$ the argument of
the tangent and cosinus can change by $2\pi$, leading to a big
difference in $\lambda (t)$ and $\xi (t)$.

A few remarks are in order. To compare with Coleman-Weinberg's
result [4], we consider the one-loop non-improved effective
potential with the standard proposal $\lambda \sim e^4$. Then one
can get from (7)
\beq
V= \frac{\lambda}{4!} \varphi^4 + \frac{3e^4\varphi^4}{64\pi^2} \ln
\frac{\varphi^2}{\mu^2} - \frac{1}{2} \xi R \varphi^2 -
\frac{1}{(8\pi)^2} e^2R \varphi^2 \ln \frac{\varphi^2}{\mu^2}.
\eeq
Choose now $\mu=<\varphi >$, where $<\varphi >$ is the vacuum
(minimum) configuration. In flat space the equation $V'(<\varphi
>)=0$ gives the precise connection between $\lambda $ and $e^4$.
However, in curved space this is not the case, and
\beq
\frac{V'(<\varphi >)}{<\varphi >} = <\varphi >^2 \left(
\frac{\lambda}{6}
+ \frac{3e^4}{32\pi^2} \right) - R \left( \xi +
\frac{e^2}{32\pi^2} \right) =0.
\eeq
Hence, we obtain the connection between  $<\varphi >$ and the
curvature corresponding to the minimum.

Since we are working in the linear curvature approximation
(supposing that the curvature correction is small), we may just impose
(without much error) the flat space condition $\lambda/6= -
3e^4/(32\pi^2)$ by hand. Then, from (9) we get
\beq
\xi = - \frac{e^2}{32\pi^2},
\eeq
and
\beq
V=  \frac{3e^4\varphi^4}{64\pi^2} \left( \ln
\frac{\varphi^2}{<\varphi >^2} - \frac{1}{2} \right) - \frac{ e^2R
\varphi^2 }{64\pi^2}\left( \ln \frac{\varphi^2}{<\varphi >^2} -
1 \right).
\eeq
Eq. (11) constitutes the generalization to curved space of the
famous Coleman-Weinberg result (Eq. (4.9) in [4]) given in
universal form. Notice that, from (11), we immediately obtain the
scalar mass which takes into account curvature effects:
\beq
m^2(s)= V''(<\varphi >) =\frac{3e^4 <\varphi >^2}{8\pi^2} -
\frac{e^2R}{32\pi^2}.
\eeq

\ni {\bf (c)  The SU(2) gauge model.}
Let us now consider the SU(2) gauge model of Ref. [9] with one
multiplet of scalars ($\varphi^a, \ a=1,2,3$) taken in the adjoint
representation of SU(2) and one or two multiplets of spinors also
taken in the adjoint representation. The Yukawa coupling acts
through only one of the spinor multiplets (see [9] for the precise
Lagrangian). This theory is asymptotically free for all coupling
constants [9] and in the case of only one spinor multiplet it is
also asymptotically conformal invariant.

The RG improved effective potential of this theory can be
calculated to be
\beq
V=  \frac{1}{4!} \varphi^4 f^4(t) k_1g^2(t)  - \frac{1}{2}
R\varphi^2 f^2(t) \left[ \frac{1}{6} + \left( \xi - \frac{1}{6}
\right) \left( 1+ \frac{a^2g^2 t}{(4 \pi)^2}\right)^{-(12-5k_1/3-
8k_2)/a^2} \right],
\eeq
where $\varphi^2 = \varphi^a\varphi^a$, $\lambda (t)=k_1 g^2(t)$,
$h^2(t)=k_2 g^2(t)$,  $g^2(t)= g^2 \left(  1+ \frac{a^2g^2 t}{(4
\pi)^2}\right)^{-1}$,  being the values of the numerical constants
$k_1$, $k_2$ and $a^2$  given in Ref. [9], and $f(t)=\left(  1+
\frac{a^2g^2 t}{(4 \pi)^2}\right)^{(6-4k_2)/a^2}$.

 In the same way one can find the RG improved effective potential
in asymptotically free GUTs (for a review and a list of
references, see [3]).
\ms

\ni {\bf (d)  The SU(5) GUT.}
Let us  now study  the RG improved potential for the SU(5) GUT [8].
In flat space this theory has been used for the discussion of
inflationary cosmology [1,2]. The one-loop potential in the linear
curvature approximation has been given in Ref. [7].

The tree-level potential has the form
\beq
V_{\mbox{tree}} = \frac{1}{4} \lambda_1 (\tr \phi^2)^2 +
\frac{1}{2} \lambda_2 \tr \phi^4  -
\frac{1}{2} \xi R \tr \phi^2,
\eeq
where $\lambda_1$ and $ \lambda_2$ are scalar couplings, and for
simplicity we suppose that there are no fermions in the theory.
Even in this case, the system of RG equations for the coupling
constants is quite complicated and can be solved only
numerically. This is why we will just consider the vector loop
contributions to the $\beta$-functions. Presumably [5] such approach
gives qualitatively the same results that would be obtained with the
inclusion of scalar couplings.

We assume that the breaking SU(5) $\rightarrow$ SU(3) x SU(2)
x U(1) has taken place. Then $\phi = \varphi$ diag $(1,1,1,-
\frac{3}{2},-\frac{3}{2})$ and
\beq
V_{\mbox{tree}} =  \frac{15}{16} (15\lambda_1
+7\lambda_2) \, \varphi^4 -
\frac{15}{4} \xi R  \varphi^2.
\eeq
Within our approach
\bea
\frac{dg(t)}{dt} &=& -\frac{5g^3(t)}{6\pi^2}, \ \  \frac{d}{dt}
 \left[ \frac{15}{4} (15\lambda_1 +7\lambda_2) \right] \equiv
\frac{d\Lambda (t)}{dt}=\frac{5625}{128\pi^2}g^4(t), \nn \\
\frac{d\xi (t)}{dt}&=&-\frac{30}{16\pi^2}\left( \xi (t)-
\frac{1}{6} \right) g^2(t), \ \ \ \gamma = -\frac{15 g^2}{16\pi^2}.
\eea

Solving Eq. (16) and substituting the result into (5) we get the RG
improved effective potential:
\beq
V = \frac{3375}{512} \left( g^2 - \frac{g^2}{1+
\frac{5g^2t}{3\pi^2}} \right) \varphi^4 f^4 (t)
- \frac{15}{4} \left[ \frac{1}{6} + \left( \xi - \frac{1}{6}
\right) \left( 1+ \frac{5g^2t}{3\pi^2} \right)^{-9/8} \right] R
\varphi^2 f^2 (t),
\eeq
where $f(t)= \left( 1+ \frac{5g^2t}{3\pi^2} \right)^{9/16}$.
\ms

This finishes our calculation of several RG improved effective
potentials corresponding to different massless gauge theories in
curved spacetime.
\ms

\ni {\bf 3. Phase transitions.}
As stated previously, it is very common nowadays to think that the
very early universe experienced several phase transitions before it
could reach its present state. It is certainly possible that a
phase transition could be induced by the resulting (very strong)
external gravitational field existing at this epoch [6,7]. We will
now discuss such possibility by using our simple (but, on the other
hand, quite general) RG improved effective potential.

We shall be concerned with first-order phase transitions where the
order parameter $\varphi$ experiences a quick change for some
critical value, $R_c$, of the curvature. Let us first consider the
$\lambda \varphi^4$ theory, in which case we can write
\beq
\frac{V}{\mu^4} = \frac{\lambda x^2}{4! \left( 1- \frac{3\lambda
\ln x}{32\pi^2} \right)}- \frac{1}{2} \epsilon y x
\left[ \frac{1}{6} + \left( \xi - \frac{1}{6} \right) \left( 1-
\frac{3\lambda \ln x}{32\pi^2} \right)^{-1/3} \right],
\eeq
where $x=\varphi^2/\mu^2$,  $y=|R|/\mu^2$, and $\epsilon =$ sgn $R$.
The critical parameters, $x_c$, $y_c$, corresponding to the first-order
phase transition are found from the conditions
\beq
V(x_c,y_c)=0, \ \ \ \ \left. \frac{\partial V}{\partial x}
\right|_{x_c,y_c} =0, \ \ \ \ \left. \frac{\partial^2 V}{\partial
x^2} \right|_{x_c,y_c} >0.
\eeq
For the one-loop effective potential the two equations (19) can be
solved analytically [7]. However, for the RG improved potential
they lead to some transcendental equations which cannot be solved
analytically. For instance, in the case of the             $\lambda
\varphi^4$-theory effective potential (18), they are
\beq
T= -\frac{\lambda}{16\pi^2} \, \frac{\frac{1}{4} +  \left( \xi -
\frac{1}{6} \right) T^{-1/3}}{\frac{1}{6} +  \left( \xi -
\frac{1}{6} \right) T^{-1/3}}, \ \ \ \ \ \ \
 \left( 16\pi^2 + \frac{\lambda}{T} \right) x + \epsilon y=0,
\eeq
where $T=1- (32\pi^2)^{-1}  3\lambda \ln x$. The simplest case,
with
an analytical solution, is $\xi = 1/6$. Here we have found no phase
transition (since $y_c \sim \pi^2 x_c$, which lies outside our
approach $x_c >> y_c$).
In the same way we can see that there is no phase transition for
$\xi = 1/6$ in the asymptotically free SU(2) model.

Consider now the RG improved potential (17). For the sake of
simplicity, let us put $f(t)=1$. (To take into account a non-zero
anomalous dimension presumably only rescales $\varphi$ [5].) \,
After choosing  $\xi = 1/6$ and solving (19), we obtain
\beq
x_c\simeq 10^2, \ \ \ \ \ \epsilon y_c \simeq g^4x_c.
\eeq
Thus, it turns out that a curvature induced phase transition is
possible in this model even in the simplest situation where $\xi =
1/6$ (there are no radiative corrections to the $R\varphi^2$-term.)

A reasonable estimation [6,7] shows that in the GUT epoch
\beq
10^{-7} \leq |y| \leq 10^{-5}.
\eeq
And for a standard choice $g^2\sim 1/3$ and $\epsilon y_c \simeq
10$, what seems to be too large and non-realistic. However, we
could argue that at the begining of inflation $g^2$ corresponds to
the {\it running} $g^2(t)$. In this case, a natural choice is [7]
$g^2\simeq 10^{-3}$ and this gives  $\epsilon y_c \simeq 10^{-4}$.
This value is already very close to the upper border of the
estimation (22), what is quite remarkable. Of course, the inclusion
of scalar loops and/or the estimation of phase transitions for
other choices of $\xi$ can also lower the value of $y_c$. This is
an interesing analysis which demands for numerical calculations.
\ms

\ni {\bf 4. Conclusions.}
To summarize, we have developed an explicit formalism for the
determination of the RG improved effective potential for massless
gauge theories in curved spacetime. We have shown the plausible
possibility of a curvature-induced phase transition taking place
for the SU(5) RG improved inflationary potential. It would be
interesting to understand the further influence of
quantum-gravitational effects in the above described picture.

In principle, this can be done, at least for
multiplicatively-renormalizable $R^2$-gravity (see [3] for a
review). Starting from the following Lagrangian (in Euclidean
notation)
\beq
L= \wt{\alpha} R^2+ \wt{\beta} R_{\mu\nu} R^{\mu\nu} + \xi R
\varphi^2 + \frac{1}{2} g^{\mu\nu} \partial_{\mu} \varphi
\partial_{\nu}\varphi + \frac{\lambda}{4!} \varphi^4,
\eeq
and forgetting for a moment about the unitarity problem (see [3]
for a  list of references concerning this point), we can employ
the background field method in order to show that the theory given
by (23) can be asymptotically free for all coupling constants [3]
(see also [10]).

The formalism developed at point 2 can be extended to the theory
(23) (see [11]) ---by just adding $\wt{\alpha}$ and  $\wt{\beta}$
to the set $\wt{g}$--- working in the background-field method and
in the one-loop approximation (then the anomalous dimension for the
background gravitational field is zero). We cannot pretend to find
the $f(t)$ of Eq. (4) explicitly either, because to this end we would
need to work out $\gamma$ in a class of gauges dependent on the
parameter. However,
as before, we can perform this calculation for the simplest gauge.
Hence, we drop the $\gamma$-dependence and put $f(t)=1$ in (4).
With all this in mind, the RG improved effective potential in the
linear curvature approximation is again given by (4). In
particular, for one of the regimes of asymptotic freedom of the
theory given by (23) [10], we get
\beq
V=  \frac{4.72}{4!} \varphi^4 \wt{\beta}^{-1} (t) - 0.03 \, R \,
\varphi^2,
\eeq
where
\[ \wt{\beta} (t) = \wt{\beta} (0)+ \frac{799}{60 (4\pi)^2} t. \]
We thus see that we actually may take into account the quantum
gravitational effects in a rather simple way.

The other interesting topic is the derivation of RG improved
effective potentials in massive theories [5,12]\footnote{It would
be meaningful also to improve the effective potential in the models
which have composite Higgs scalars in curved space [14].}. There
has been some recent activity in this direction [13], readdressing
the question in flat space. It is our impression that quantum field
theory in curved spacetime can really help to answer some questions
about the RG improved effective potential for massive theories even
in flat space. Work along this line is in progress. Notice,
finally, that the generalization to the multiscale RG can be
carried out as in Ref. [15] in flat space.
%\vspace{5mm}

\newpage

\ni{\bf Acknowledgments.}

S.D.O. wishes to thank  JSPS (Japan) for financial support and the
Particle Group at Hiroshima University for kind hospitality, and
E.E.  the DGICYT (Spain), for financial help through research
project PB90-0022. We are very grateful to T. Muta for his interest
in this work and for helpful discussions.
\bs

\ni{\bf Appendix.}
We shall here study the behaviour of the RG improved effective
potential (18) in the $\lambda \varphi^4$ model. We shall also
compare it with that of the one-loop effective potential for the same
theory, which is
\beq
\frac{V_{\mbox{one-loop}}}{\mu^4} = \frac{\lambda x^2}{4!} \left( 1+
\frac{3\lambda \ln
x}{32\pi^2} \right)- \frac{1}{2} \epsilon y x
\left[ \xi + \left( \xi - \frac{1}{6} \right)
\frac{\lambda \ln x}{32\pi^2} \right].
\eeq
(This is just an expansion of (18), taking $\lambda <<1$ and
$|\lambda \ln x | <<1$.) \, A very important difference between (18) and
(25) lies in the fact that while the RG improved effective potential
(18)
has a pole for $x=x_p\equiv \exp [32 \pi^2 /(3 \lambda)]$, the one-loop
effective potential (25) is a smooth function for the whole range of
values of $x$ and $y$. Notice however, in particular, that expression
(18) is finite for all negative values of $\ln x$. The fact that the
$\lambda \varphi^4$ potential (25)
exists for {\it all} values of $\varphi$ makes this expression obviously
wrong.

We perform the usual analysis of extrema of (18) and (25). Calling
$V(x,y)$ the potential in each case, we shall look for critical points
$(x_c, y_c)$ defined by the simultaneous conditions (19).
The first two equations (19) yield, for the RG improved action (18),
\beq
y_c=\frac{\epsilon \lambda x_c}{ 2 u \left[ 1+(6\xi-1)
u^{-1/3}\right]}, \ \
 \frac{32\pi^2}{3\lambda}\, u -\frac{1}{3\left[ 1+ (6\xi
-1)u^{1/3}\right]} +1=0,
\ \ u\equiv 1- \frac{3\lambda \ln x_c}{32\pi^2},
\label{ce18}
\eeq
and, for the one-loop action (25),
\beq
y_c=\frac{\epsilon \lambda (1+3v) x_c}{12\left[ \xi + \left( \xi -
\frac{1}{6} \right) v\right]}, \ \
 \frac{32\pi^2}{\lambda} \left( v+\frac{1}{3}\right) -
\frac{\lambda}{96\pi^2} \left( \xi
- \frac{1}{6} \right) \frac{1+3v}{\xi + \left( \xi - \frac{1}{6} \right)
v}=0, \ \
 v \equiv \frac{\lambda \ln x_c}{ 32\pi^2}.
\label{ce25}
\eeq

 The following models, which are particularly interesting for different
reasons, will be considered in further detail. \ms

\ni {\bf (a)  Chaotic inflationary model.}
We  consider first the potentials (18) and (25) for  $\lambda =
10^{- 13}$ and $\xi =0$, both for positive $\epsilon =1$ and negative
$\epsilon =-1$ curvature. The results can be summarized as follows.
For the RG improved effective action, a critical value appears, which
lies close to the pole
\beq
x_c=\exp \left(- \frac{2}{3} \, 10^{-15} \right) \, x_p, \ \ \ \
y_c=- \epsilon \, 10^{-3} \, x_c.
\eeq
Moreover, this point is a minimum of (18) (as are all the similar points
obtained below). That is, all three equations (19) are indeed
satisfied.
 On the contrary, the
one-loop effective action (25) does
not yield any phase transition, the solutions of Eqs. (27) being
$x_c=0$, $y_c=0$, and
$x_c=1$, $y_c= \epsilon \, 10^{17} \, x_c$.
\ms

\ni {\bf (b)  Variable Planck-mass model.}
It has been considered in [17]. For  $\lambda$ we take a typical
value corresponding to particle physics models, e.g. $\lambda = 0.05$.
For
$\xi$ we choose two different values: (i) $\xi = - 10^4$
(which actually corresponds to ref.[17])
 and (ii)  $\xi
= 1/6$, respectively.

 In case (i), the critical point corresponding to
(18) is obtained for
\beq
x_c= e^{2/3} \, x_p, \ \ \ y_c= -5 \cdot 10^{-5} \, x_c,
\eeq
both for positive and for negative curvature. For the one-loop action,
the only solution is again the trivial one $x_c=y_c=0$.

 In  case (ii), the critical point for the RG improved action
(18) is at
\beq
x_c= e \, x_p, \ \ \ y_c= -\epsilon \, 50 \, x_c,
\eeq
which is not consistent with our approximation $x_c>>y_c$.
For the one-loop effective action (25),
 $x_c=y_c=0$ is again the only solution of (27).

\newpage
\renewcommand
\baselinestretch{1.1}
\end{document}